# Stochastic Optimal Operation of the VSC-MTDC System with FACTS Devices to Integrate Wind Power

Zhao Yuan, *Member, IEEE*, Mohammad Reza Hesamzadeh, *Member, IEEE*, Sonja Wogrin, *Member, IEEE*, Mohamadreza Baradar

*Abstract*—This paper proposes to use stochastic conic programming to address the challenge of large-scale wind power integration to the power system. Multiple wind farms are connected through the voltage source converter (VSC) based multi-terminal DC (VSC-MTDC) system to the power network supported by the Flexible AC Transmission System (FACTS). The optimal operation of the power network incorporating the VSC-MTDC system and FACTS devices is formulated in a stochastic conic programming framework accounting the uncertainties of the wind power generation. A methodology to generate representative scenarios of power generations from the wind farms is proposed using wind speed measurements and wind turbine models. The nonconvex transmission network constraints including the VSC-MTDC system and FACTS devices are convexified through the proposed second-order cone AC optimal power flow model (SOC-ACOPF) that can be solved to the global optimality using interior point method. In order to tackle the computational challenge due to the large number of wind power scenarios, a modified Benders decomposition algorithm (M-BDA) accelerated by parallel computation is proposed. The energy dispatch of conventional power generators is formulated as the master problem of M-BDA. The wind power dispatch problem is formulated and decomposed by scenarios as the subproblems of M-BDA which can be solved in parallel. We make full use of the multi-CPU computer architecture by balancing the number of wind power scenarios of all the formulated subproblems in M-BDA. Numerical results for up to 50000 wind power scenarios show that the proposed M-BDA approach to solve stochastic SOC-ACOPF outperforms the traditional single-stage (without decomposition) solution approach in both convergence capability and computational efficiency. The feasibility performance of the proposed stochastic SOC-ACOPF model is also demonstrated numerically.

## Nomenclature

**Sets:**

| | |
|---|---|
| $i_{AC}, i_{DC}$ | AC and DC Buses. |
| $l_{AC}$ | AC Lines. |
| $l_{DCmono}$ | Mono-polar VSC-MTDC Lines. |
| $l_{DCbi}$ | Bi-polar VSC-MTDC Lines. |
| $J$ | Scenarios. |
| $E$ | Wind Farms. |
| $K$ | Iterations. |

Zhao Yuan is with the Electrical Power Systems Laboratory (EPS-Lab), Department of Electrical and Computer Engineering, University of Iceland, Iceland, Iceland: zhaoyuan@hi.is. Mohammad Reza Hesamzadeh is with KTH Royal Institute of Technology. Sonja Wogrin is with Comillas Pontifical University. Mohamadreza Baradar is with Vattenfall.

**Parameters:**

| | |
|---|---|
| $A^{+}_{i,l}, A_{i,l}$ | Bus to Line Incidence Matrix. |
| $R_l, X_l$ | Resistance and Reactance of Line $l$. |
| $R_T, X_T$ | Transformer Resistance and Reactance |
| $R_{Cse}, R_{Csh}$ | VSC Station Serial and Shunt Resistance |
| $G_i, B_i$ | Shunt Conductance and Susceptance at Bus $i$. |
| $K_l$ | Squared Power Capacity of Line $l$. |
| $p^{Min}_i, p^{Max}_i$ | Lower and Upper Bound of $p_i$. |
| $q^{Min}_i, q^{Min}_i$ | Lower and Upper Bound of $q_i$. |
| $D_i, Q_i$ | Active and Reactive Power Demand at Bus $i$. |

**Variables:**

| | |
|---|---|
| $p_i, q_i$ | Active and Reactive Power Generation (thermal generator). |
| $p_{e_{i,j}}, q_{e_{i,j}}$ | Active and Reactive Power Generation (wind). |
| $p_{s_l}, q_{s_l}$ | Active and Reactive Power Injection of Line $l$. |
| $p_{loss_l}, q_{loss_l}$ | Active and Reactive Power Loss of Line $l$. |
| $v_i, V_i$ | Voltage Magnitude (lower case) and Its Square (upper case) at Bus $i$. |
| $v_{s_l}, V_{s_l}$ | Voltage Magnitude (lower case) and Its Square (upper case) at the Sending End of Line $l$. |
| $v_{r_l}, V_{r_l}$ | Voltage Magnitude (lower case) and Its Square (upper case) at the Receiving End of Line $l$. |
| $\theta_l$ | Voltage Phase Angle Difference of Line $l$. |
| $\theta_{s_l}, \theta_{r_l}$ | Phase Angle at the Sending and Receiving End of Line $l$. |
| $I_{s_l}$ | Current at the Sending End of Line $l$. |

## I. Introduction

### A. Motivation

**P**Ower systems are gaining many operational flexibilities and benefits from power electronics based equipments such as Voltage Source Converter (VSC) based multi-terminal DC (VSC-MTDC) system and Flexible AC Transmission System (FACTS) [1], [2]. Especially, VSC-MTDC and FACTS are playing important roles in integrating intermittent renewable energy resources [3], [4]. On the one hand, VSC-MTDC and FACTS can be applied to remove congestion in heavily loaded power systems [5]. On the other hand, FACTS devices can be used to enhance the transfer capacity of transmission network [6]. Claus et al. [7] point out that flexible hybrid AC-DC systems are inevitable in the future smart grid.

Other authors have investigated the impact of FACTS devices on the system loadability enhancement [8], [9] and studied the optimal power flow (OPF) problem with minimal transmission losses using the multi-terminal high-voltage DC (HVDC) systems embedded in a mesh power grid [10]. Ref. [11] studies the impacts of back-to-back VSC-type HVDC systems on the power loss of mesh network and uses a load flow program solved by Newton-Raphson method to evaluate the network power loss. Accordingly, the operation points of HVDC are fixed before running the load flow program and the loss is calculated. In [12] the authors study the impact of a Unified Power Flow Controller on the optimal reactive power dispatch problem. The importance of VSC-MTDC systems and FACTS is further emphasized by [13], which investigates the benefits of FACTS devices in the IEEE 57-Bus test system with high penetration of wind power. After installing FACTS devices, the total savings of the net present value (NPV) over 20 years amount to \$49.45 million based on the probabilistic modeling of the load growth.

Given the forecasted power load and renewable energy generation, for a power network (with FACTS devices) connected to wind farms through VSC-MTDC system, what is the optimal operating point (voltage, power flow, power dispatch) that the power system operator should set for the power network? The operator should minimize the economic cost or maximize the engineering efficiency of the power network while taking all the operational constrains into account. The economic benefit or engineering gain of solving this problem is huge considering the scale of power network, the volume of transmitted energy, and growing penetration of renewable energy resources. The major challenges of solving this operating problem (focus of this paper) lie in the complexity of the formulated OPF model (including power network constraints) and the uncertainty of renewable energy (wind) generation. To deal with the complexity of the OPF model, we propose a convexification approach (SOC-ACOPF) to reformulate the original nonconvex OPF model to a convex one which shows efficient computational performance in terms of global optimality. For the uncertainty of renewable energy generation, our solution is based on stochastic programming using scenarios to represent the wind power generation forecast. The computational challenge, due to large number of wind power scenarios, is then solved by our proposed M-BDA capable of decomposing and parallelizing the stochastic SOC-ACOPF model.

### B. Literature Survey

The most relevant literature regarding the optimal operation of power network including VSC-MTDC system and FACTS devices are generally modeled in an OPF problem. Ref. [14] proposes a decentralized DC OPF algorithm based on the projected gradient method to minimize both generation cost and discomfort cost caused by demand responses. The application of oblivious routing algorithm in power routing for clusters of DC micro-grids is investigated by [15]. Compared with DC OPF, AC OPF can give more accurate solutions. The AC OPF consists of a set of nonconvex constraints and is often solved using Newton-Raphson methods [16], [17], which cannot guarantee to find the global optimum. Alternatively, authors in [18] propose a predictor-corrector modified barrier approach to address the nonconvex transmission network constraints. Using non-deterministic optimization search techniques, [19] and [20] minimize the network losses in an AC OPF setting. Because of the nonconvex nature of the AC OPF model, heuristic algorithms including gravitational search [21], artificial bee colony algorithm [22] and genetic algorithm [23] are often deployed to solve the AC OPF model. Ref. [24] uses differential evolution to solve the formulated multi-objective AC OPF model taking some FACTS devices into account. Many useful modelling approaches or solution techniques from the operations research area such as the linear programming approximation in [25], the second-order cone programming approaches [26] and the branch-and-bound algorithms based on Lagrangian duality analysis in [27] have also been proposed. These approaches either simplify or relax the power flow constraints [25], [26] or require very large computational efforts [27]. A relaxed convex OPF model of AC-DC grids based on semi-definite programming (SDP) is formulated in [28] to jointly minimize the generation cost, power loss of the lines and converters. Using a barrier term in the objective function, the voltage variations caused by random renewable energy generations can be reduced in the SDP based AC OPF model [29]. The computational limits of SDP are shown in [30]. Efficient algorithms for solving SDP-based AC OPF model remain to be found [31]. Considering the computational advantage of the second-order cone programming (SOCP) based AC OPF model [32]–[34] over the SDP-based AC OPF model, we use the SOCP-based AC OPF model in this paper.

Another challenge of operating the power network including VSC-MTDC system and FACTS is the stochastic nature of renewable energy resources such as wind power. Stochastic programming has been proved to be a powerful approach to deal with the uncertainties in power system operations in [35], [36] To consider the uncertainty of wind power, authors in [37] propose a scenario generation method that accounts for wind power forecast errors and fluctuation distributions. The interdependence structure of prediction errors in generating wind scenarios is the focus of [38]. Ref. [39] describes a local search algorithm to determine the operation points of two HVDC systems connecting Jeju Island to the Korean Peninsula. The potential of demand responses to deal with the uncertainty of wind power are investigated by [40] using a stochastic programming approach. Second-order moment based distributional uncertainty set is used in [41] to capture the correlation between wind power output and transmission line ratings. To ensure small possibility of violating the transmission line capacity due to uncertain renewable energy generations, a data driven distributional robust chance constrained OPF is formulated in [42]. A SOCP reformulation of the OPF model is also proposed in [42] to make the model easier to be solved. Ref. [43] formulates a stochastic nonconvex multi-period OPF model to integrate wind generation through HVDC system. The HVDC type considered in this work is line-commutated converters based HVDC (LCC-HVDC). The power generations of thermal (conventional) generators are taken as the first-stage variables (fixed for all wind power scenarios) and the second stage-variables (determined according to each wind power scenario) include the wind power generations and HVDC operation points. Although [43] models multiple wind farms, the total number of considered wind power scenarios is limited to twelve. Ref. [44] minimizes the wind power spillage by formulating a stochastic non-convex OPF model including FACTS devices. Similarly, the power outputs of thermal generators are fixed for all wind power scenarios. The second-stage variables include the wind power outputs and the control variables of FACTS devices. We use the same approach in [43], [44] to divide first-stage and second-stage variables in the stochastic programming model of this paper.

Considering the uncertainty of the wind power generation in the VSC-MTDC system and FACTS leads to a severe computational challenge. For example, if we consider the case of Sweden, there are currently around 38 wind farms [45]. If each wind farm were operated by an individual company, and only two possible scenarios for wind power outputs are taken into account, the total number of scenarios to be analyzed by the transmission system operator (TSO) would be $2^{38}$. The scale of the formulated OPF problem grows exponentially. This large-scale computational challenge is considered in [46], where parallel computation is used to solve a multi-objective security constrained OPF model considering 1000 wind power scenarios. The parallel computation carried out in MATLAB takes more than two hours to find the solutions in [46]. As an improvement, the maximum number of wind power scenarios addressed in this paper is up to 50000. [47] proposes a bounding-based method to solve the formulated large-scale stochastic optimization problem where the expected wind power generation cost is minimized in a multi-area power system. Benders decomposition is used to find the upper bound of the expected power generation cost by solving the linear program in [47]. [48] proposes to modify and use Benders decomposition to accelerate solving the SOC-ACOPF model by partitioning the power network. Coordinating the operations of power transmission system and distribution system by using the formulated Benders decomposition is proposed in [49]. These references show the strong capability of Benders decomposition in solving large-scale power system optimization problems [47]–[49].

### C. Contribution and Organization of Current Article

The contributions of this paper are:

1) We propose a convex SOC-ACOPF model that can be solved efficiently by the interior point method (IPM). Compared with the SOCP-based AC OPF model in [32], we explicitly include voltage phase angle variables in the model and thus these solutions can be obtained directly by solving our model. Furthermore, we extend the model to include both mono-polar and bi-polar VSC-MTDC system. A computational comparison shows better AC feasibility performance of the proposed SOC-ACOPF model over the widely used DC OPF model.

2) A methodology to generate representative wind power scenarios of multiple wind farms based on wind speed measurements and wind turbine models is proposed;

3) Based on the proposed SOC-ACOPF model and wind power scenario generation methodology, the optimal operation of power network including VSC-MTDC system and FACTS devices is formulated as a stochastic SOC-ACOPF model to consider the uncertainties of wind power. We also give the expressions for the scale of the formulated stochastic SOC-ACOPF model in terms of number of variables and constraints.

4) The large-scale computational challenge of up to 50000 wind power scenarios is addressed by the proposed M-BDA and parallel computation approaches; A fundamental reason for the improved performance is that by using M-BDA, we actually reduce the dimension of the Hessian and Jacobian matrices during the iteration of the deployed IPM-based solver (MOSEK). The efficient design of the parallel computation structure to balance the number of scenarios in each formulated subproblem of M-BDA in GAMS platform is another reason for the improved performance.

The rest of this paper is organized as follows. Section II formulates the exact nonlinear model of the hybrid AC-DC power network including FACTS devices. Section III proposes the convexification and approximation methods for the AC OPF model. Section IV introduces the wind power scenario generation methodology. Section V presents the stochastic SOC-ACOPF model. Section VI describes the parallel computation structure using GAMS platform based on the proposed M-BDA of the stochastic SOC-ACOPF model. Section VII presents the numerical results and discusses the efficiency of the proposed approaches. Finally, Section VIII concludes this paper. The structure of all the aforementioned Sections in this paper is illustrated in Fig. 1 where we plot the key building blocks and their relationship. The wind power scenario generation methodology is based on sufficient wind data acquisition and probabilistic modelling. The SOC-ACOPF model is based on the AC OPF model including the VSC-MTDC system and FACTS devices. A stochastic SOC-ACOPF model is formulated by using the wind power scenario generation methodology and SOC-ACOPF model. This stochastic SOC-ACOPF model are solved by using three approaches: single-stage solution approach, serial M-BDA and parallel M-BDA. We also compare the performance of these three solution approaches to demonstrate the efficiency of parallel M-BDA.

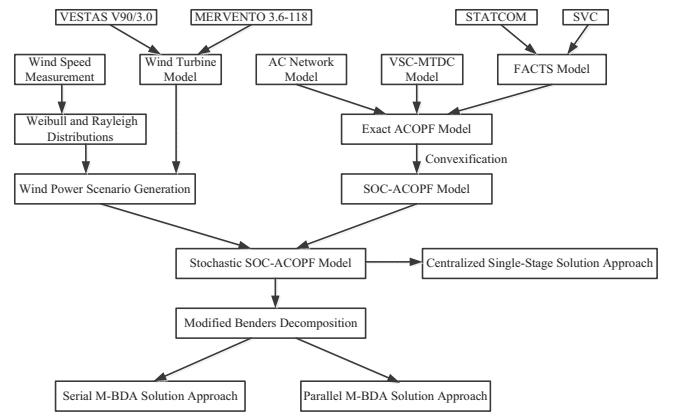

Fig. 1. Structure of the contents.

## II. Hybrid AC-DC network and FACTS models

### A. The AC network

We use variables $p_{s_l}, q_{s_l}$ to represent active and reactive power flow at the sending end of line $l$, respectively. Variables $p_{loss_l}, q_{loss_l}$ represent active and reactive power loss of line $l$. $i_{AC}$ is defined as the set of AC buses. $l_{AC}$ is the set of AC lines (both ends of the line are AC buses). The derived constraints representing the AC network are listed as follows [32], [50]–[52]. Note that since the variables $p_{s_l}, q_{s_l}$ are different from [32], some constraints are accordingly different. The subscript $_s$ in $p_{s_l}, q_{s_l}, v_{s_l}$ is not an index but only to imply the meaning of sending end. The subscript $_r$ in $v_{r_l}$ is not an index but only to imply the meaning of receiving end. Similar reasoning holds for the subscript $_{loss}$ in $p_{loss_l}, q_{loss_l}$. To distinguish AC network from the DC network, we have specified the sets $i_{AC}, l_{AC}$ for all the indices at the end of the corresponding constraints.

$$p_i - D_i = \sum_{l \in l_{AC}} (A_{i,l}^+ p_{s_l} - A_{i,l}^- p_{loss_l}) + G_i v_i^2, \ \forall i \in i_{AC} \tag{1a}$$

$$q_i - Q_i = \sum_{l \in l_{AC}} (A_{i,l}^+ q_{s_l} - A_{i,l}^- q_{loss_l}) - B_i v_i^2, \ \forall i \in i_{AC} \tag{1b}$$

$$p_{loss_l} = \frac{p_{s_l}^2 + q_{s_l}^2}{v_{s_l}^2} R_l, \ \forall l \in l_{AC} \tag{1c}$$

$$q_{loss_l} = \frac{p_{s_l}^2 + q_{s_l}^2}{v_{s_l}^2} X_l, \ \forall l \in l_{AC} \tag{1d}$$

$$v_{s_l}^2 - v_{r_l}^2 = 2(R_l p_{s_l} + X_l q_{s_l}) - R_l p_{loss_l} - X_l q_{loss_l}, \forall l \in l_{AC} \tag{1e}$$

$$v_{s_l} v_{r_l} \sin \theta_l = X_l p_{s_l} - R_l q_{s_l}, \ \forall l \in l_{AC} \tag{1f}$$

where the variable $v_i$ is the voltage magnitude at AC bus $i$. The variables $v_{s_l}$ and $v_{r_l}$ are the voltage magnitudes at the sending and receiving ends of AC line $l$, respectively. $A_{i,l}^+$ and $A_{i,l}^-$ are parameters representing network topology. $A_{i,l}^+ = A_{i,l}^- = 1$ if $i$ is the sending end of line $l$ and $A_{i,l}^+ = -1, A_{i,l}^- = 0$ if $i$ is the receiving end of line $l$. The default direction of each line is defined by the topology of the network. $\theta_l = \theta_{s_l} - \theta_{r_l}$ is the voltage phase angle difference between the sending and receiving ends of AC line $l$. $p_i$ and $q_i$ are the active and reactive power generation at each AC bus. Parameters $D_i$ and $Q_i$ are active and reactive power loads at AC bus $i$. $R_l, X_l$ are the resistance and reactance of line $l$. $G_i, B_i$ represent the shunt conductance and capacitance at AC bus $i$. The nonlinear constraints (1a) and (1b) are active and reactive power balance equations for AC buses; Nonlinear constraints (1c)-(1e) represent the active and reactive power losses, and voltage drop constraint for each AC line; Finally, (1f) which is nonlinear, is associated with phase angle difference for each AC line.

### B. The VSC-MTDC system

*1) The DC side:* The equivalent circuit in Fig. 2 is used to derive the constraints for the AC-DC grid and the VSC stations. In Fig. 2, PCC is short for point of common coupling. PC is the AC bus coupling with the DC bus of VSC. The VSC is connected with the PCC through a coupling transformer. $R_{Cse}, R_{Csh}$ are resistors to represent the ohmic losses of VSC.

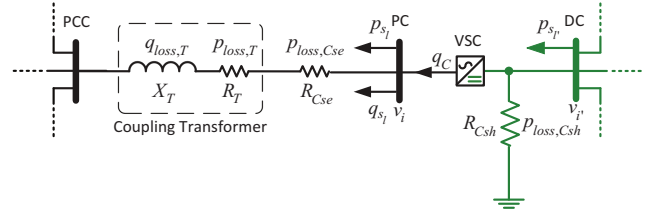

Fig. 2. The equivalent circuit of VSC-MTDC system.

We define $i_{DC}$ as the set of DC buses, and $l_{DC}$ as the set of DC lines connected to the VSC. In order to distinguish the DC network from the AC network, we have specified the sets $i_{DC}, l_{DC}$ for all the indexes at the end of each constraint. The active power balance equation at each DC bus can be written as:

$$p_i = \sum_{l \in l_{DC}} (A_{i,l}^+ p_{s_l} - A_{i,l}^- p_{loss_l}) + G_i v_i^2, \ \forall i \in i_{DC} \tag{2a}$$

where $p_i$ is the injected active power at DC bus $i$.

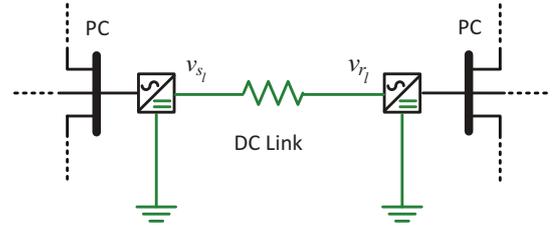

Fig. 3. The monopole MTDC connection.

For mono-polar DC connections shown in Fig. 3, we define $l_{DCmono} \in l_{DC}$ as the set of monopole DC links. The power loss of each monopole DC line is formulated as:

$$p_{loss_l} = \frac{p_{s_l}^2}{v_{s_l}^2} R_l, \ \forall l \in l_{DCmono} \tag{2b}$$

The voltage square drop across each monopole DC link is:

$$\begin{aligned} v_{s_l}^2 - v_{r_l}^2 &= v_{s_l}^2 - (v_{s_l} - \Delta v_l)^2 \\ &= v_{s_l}^2 - v_{s_l}^2 - \Delta v_l^2 + 2 v_{s_l} \Delta v_l \\ &= 2 v_{s_l} I_{s_l} R_l - (I_{s_l} R_l)^2 \\ &= 2 p_{s_l} R_l - p_{loss_l} R_l, \ \forall l \in l_{DCmono} \end{aligned} \tag{2c}$$

where $\Delta v_l = v_{s_l} - v_{r_l}$ is the voltage drop of the DC link. $I_{s_l}$ is the current of the DC link. We make use of $\Delta v_l = I_{s_l} R_l$, $p_{s_l} = v_{s_l} I_{s_l}$ and $p_{loss_l} = I_{s_l}^2 R_l$ in the derivations of (2c). Although we use $\Delta v_l, I_{s_l}$ to derive (2c), these variables are not included in our proposed SOC-ACOPF model since the solutions of these variables can be recovered from those of $p_{s_l}, v_{s_l}, v_{r_l}$.

For bipolar DC connections shown in Fig. 4, we define $l_{DCbi} \in l_{DC}$ as the set of bi-polar DC line. The power loss

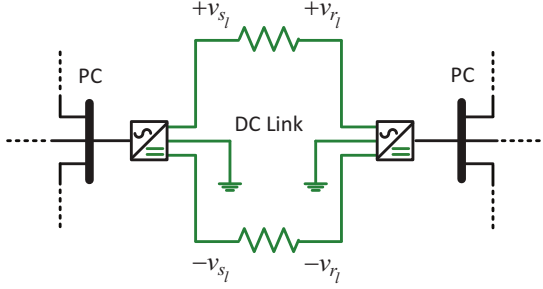

Fig. 4. The bipolar MTDC connection.

of each bipolar DC line is formulated as (Note here $R_l$ is the total resistance of the bi-polar DC link):

$$p_{loss_l} = \frac{p_{s_l}^2}{4v_{s_l}^2}R_l, \ \forall l \in l_{DCbi} \tag{2d}$$

Similarly, the voltage drop across each bi-polar DC link is:

$$v_{s_l}^2 - v_{r_l}^2 = p_{s_l}R_l - p_{loss_l}R_l, \ \forall l \in l_{DCbi} \tag{2e}$$

Equations (2d)-(2e) are valid because $p_{s_l} = 2v_{s_l}I_{s_l}$ for bipolar DC links. The relationship between the AC and DC sides are represented using voltage and power couplings. We define $i_{PC}$ as the set of AC buses coupling with the DC buses of VSC. As it is shown in Fig. 2, the DC side voltage at each station is related to its AC side voltage:

$$v_i = \frac{v_i'}{2\sqrt{2}}m_{i,i'}, \ \forall i \in i_{PC}, \forall i' \in i_{DC} \tag{2f}$$

where $m_{i,i'}$ indicates the modulation index for the converter between the bus $i$ and $i'$. In practice the modulation index is bounded as $0.5 \leq m_{i,i'} \leq 1$. Several steady state models for converter loss have been used in the literature [16], [53]. In this paper, we consider the loss model proposed in [53] where the converter losses are represented by a series resistor ($R_{Cse}$) on the AC side and a shunt resistor ($R_{Csh}$) on the DC side (See Fig. 2). The loss of the shunt resistor $R_{Csh}$ can be calculated as:

$$p_{loss,Csh} = \frac{v_{i'}^2}{R_{Csh}}, \ \forall i' \in i_{DC} \tag{2g}$$

*2) The AC side:* Define $l_{VSC}$ as the converter line (from the DC bus to the PC bus) and $l_{PC}$ as the AC line coupling with the converter (from the PC bus to the PCC bus). The sending end power of the converter line can be obtained by:

$$p_{s_l} = p_{s_{l'}} - p_{loss,Csh}, \ \forall l \in l_{PC}, \forall l' \in l_{VSC} \tag{2h}$$

$$q_{s_l} = q_C, \ \forall l \in l_{PC}, \forall l' \in l_{VSC} \tag{2i}$$

where $q_C$ is the reactive power output of the voltage converter station in the VSC-MTDC system (the subscript $_C$ is an index of voltage converters). The constraints for the AC side of each converter station are derived using the equivalent circuit shown in Fig. 2. As it is shown, coupling transformer and phase reactor can be simply modeled as an AC line. Therefore constraints (1e) and (1f) together with following equations can be applied to each VSC station. The power loss of coupling transformer $p_{loss,T}$ $q_{loss,T}$, and AC side power loss of VSC

$q_{loss,Cse}$ are obtained in a similar way as (1c) and (1d) in the following equations:

$$p_{loss,T} = \frac{p_{s_l}^2 + q_{s_l}^2}{v_{s_l}^2}R_T, \ \forall l \in l_{PC} \tag{2j}$$

$$q_{loss,T} = \frac{p_{s_l}^2 + q_{s_l}^2}{v_{s_l}^2}X_T, \ \forall l \in l_{PC} \tag{2k}$$

$$p_{loss,Cse} = \frac{p_{s_l}^2 + q_{s_l}^2}{v_{s_l}^2}R_{Cse}, \ \forall l \in l_{PC} \tag{2l}$$

where $R_T$ and $X_T$ are resistance and reactance of the coupling transformer.

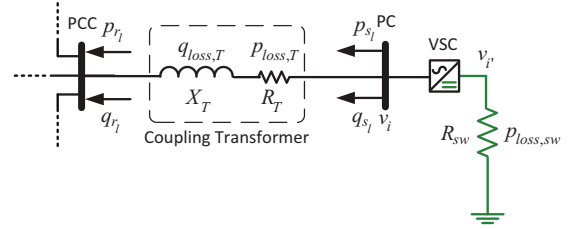

Fig. 5. The equivalent circuit of STATCOM

### C. FACTS devices

The FACTS devices considered in this paper include static synchronous compensator (STATCOM) and static VAR compensator (SVC).

*1) STATCOM:* The coupling transformer between the VSC in STATCOM and connecting point to the network can be modeled as an AC line (See Fig. 5). Similar as the MTDC model, $l_{PC}$ is defined as the coupling transformer line from the PC bus to the PCC bus (the subscript $_{PC}$ is an index of the coupling transformer lines). The STATCOM can be modeled using the following set of equations.

$$q_{r_l} = q_{s_l} - q_{loss,T}, \ \forall l \in l_{PC} \tag{3a}$$

$$p_{r_l} = p_{s_l} - p_{loss,T}, \ \forall l \in l_{PC} \tag{3b}$$

$$v_{s_l}^2 - v_{r_l}^2 = 2R_T p_{s_l} + 2X_T q_{s_l} - R_T p_{loss,T} - X_T q_{loss,T},$$
$$\forall l \in l_{PC} \tag{3c}$$

The losses of STATCOM are modeled by:

$$q_{loss,T} = \frac{p_{s_l}^2 + q_{s_l}^2}{v_{s_l}^2}X_T, \ \forall l \in l_{PC} \tag{3d}$$

$$p_{loss,T} = \frac{p_{s_l}^2 + q_{s_l}^2}{v_{s_l}^2}R_T, \ \forall l \in l_{PC} \tag{3e}$$

A shunt resistor in the DC side of the converter is used to model the switching power losses. The DC side voltage of STATCOM is related to the AC side voltage as follows:

$$v_i = \frac{v_{i'}}{2\sqrt{2}}m_{i,i'}, \ \forall i \in i_{PC}, \forall i' \in i_{DC} \tag{3f}$$

The switching power losses are calculated by:

$$p_{loss,sw} = \frac{v_{i'}^2}{R_{sw}}, \ \forall i' \in i_{DC} \tag{3g}$$

*2) SVC:* The equivalent circuit of the SVC is shown in Fig. 6. We define $l_{SVC}$ as the set of SVC lines (from the SVC device to the connected AC bus. the subscript $_{SVC}$ is an index of the SVC devices). The SVC is modeled as a variable susceptance with upper and lower bounds $B_l^{Max}, B_l^{Min}$ [54]. Accordingly, the SVC constraints are:

$$q_{s_l} = -b_l v_{s_l}^2, \ \forall l \in l_{SVC} \tag{3h}$$

$$B_l^{Min} \leq b_l \leq B_l^{Max}, \ \forall l \in l_{SVC} \tag{3i}$$

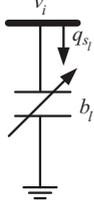

Fig. 6. The equivalent circuit of SVC.

## III. Convexification and approximation

The convexfication and approximation of the hybrid AC-DC network and FACTS models are derived in this section. The convexification is based on second-order cone programming. The approximation is based on small voltage phase angle difference assumption and voltage magnitude assumption for power transmission network.

### A. Formulation

We define a new variable $V_i = v_i^2$ for both AC and DC bus voltage magnitudes to linearize some constraints related with the voltage. Accordingly, $V_{s_l} = v_{s_l}^2, V_{r_l} = v_{r_l}^2$. The voltage solutions can be recovered by $v_i = \sqrt{V_i}$ from the solutions of $V_i$. Note we do not include the constraints $V_i = v_i^2, V_{s_l} = v_{s_l}^2, V_{r_l} = v_{r_l}^2$ in the model in order to avoid non-convexity. As long as the voltage solutions can be eventually recovered, it is not necessary to include these constraints in the model. Using these new variables, (1a)-(1b) can be written as:

$$p_i - D_i = \sum_{l \in l_{AC}} (A_{i,l}^+ p_{s_l} - V_{i,l}^- p_{loss_l}) + G_i V_i, \ \forall i \in i_{AC} \tag{4a}$$

$$q_i - Q_i = \sum_{l \in l_{AC}} (A_{i,l}^+ q_{s_l} - A_{i,l}^- q_{loss_l}) - B_i V_i, \ \forall i \in i_{AC} \tag{4b}$$

Constraint (1e) is reformulated as:

$$V_{s_l} - V_{r_l} = 2R_l p_{s_l} + 2X_l q_{s_l} - R_l p_{loss_l} - X_l q_{loss_{l,j}}, \\ \forall l \in l_{AC} \tag{4c}$$

Constraints (2c)-(2e) are rewritten as:

$$V_{s_l} - V_{r_l} = 2p_{s_l} R_l - p_{loss_l} R_l, \ \forall l \in l_{DCmono} \tag{4d}$$

$$V_{s_l} - V_{r_l} = p_{s_l} R_l - p_{loss_l} R_l, \ \forall l \in l_{DCbi} \tag{4e}$$

Constraint (3c) can be rewritten as:

$$V_{s_l} - V_{r_l} = 2R_T p_{s_l} + 2X_T q_{s_l} - R_T p_{loss,T} - X_T q_{loss,T}, \\ \forall l \in l_{PC} \tag{4f}$$

Constraints (2g) and (3g) are rewritten as:

$$p_{loss,Csh} = \frac{V_i}{R_{Csh}}, \ \forall i \in i_{DC} \tag{4g}$$

$$p_{loss,sw} = \frac{V_i}{R_{sw}}, \ \forall i \in i_{DC} \tag{4h}$$

Also constraints (3h)-(3i) can be rewritten as:

$$-B_l^{Max} V_{s_l} \leq q_{s_l} \leq -B_l^{Min} V_{s_l}, \ \forall l \in l_{SVC} \tag{4i}$$

Squaring both sides of (2f) and (3f) gives us (using $V_i = v_i^2$):

$$V_i = \frac{V_{i'}}{8} m_{i,i'}^2, \ \forall i \in i_{PC}, \forall i' \in i_{DC} \tag{4j}$$

Considering:

$$M_{i,i'}^{Min} \leq m_{i,i'}^2 \leq M_{i,i'}^{Max} \tag{4k}$$

One can obtain the following linear constraint equivalent to (4j)-(4k):

$$\frac{8V_i}{M_{i,i'}^{Max}} \leq V_{i'} \leq \frac{8V_i}{M_{i,i'}^{Min}}, \ \forall i \in i_{PC}, \forall i' \in i_{DC} \tag{4l}$$

We use two approximations $sin\theta_l \approx \theta_l$ and $v_{s_l} v_{r_l} \approx 1$ (per unit value [*p.u.*]) for (1f). These assumptions are valid for transmission network under normal operations. Note that these assumptions are only used to linearize constraint (1f). This means our model approximates fewer AC constraints than the widely used DC OPF model and thus the feasibility property (satisfying the original AC OPF constraints) of SOC-ACOPF model is better than the DC OPF model.

$$\theta_l = X_l p_{s_l} - R_l q_{s_l}, \ \forall l \in l_{AC} \tag{4m}$$

The last step is to handle the nonlinear constraints associated with power loss. First we obtain a linear relation between active power and reactive power loss:

$$p_{loss_l} X_l = q_{loss_l} R_l, \ \forall l \in l_{AC} \tag{4n}$$

We replace all equalities in the form of (1c), (2b), (2d) and (3d) with the following inequalities:

$$q_{loss_l}^{Max} \geq q_{loss_l} \geq \frac{p_{s_l}^2 + q_{s_l}^2}{V_{s_l}} X_l, \ \forall l \in l_{AC} \tag{4o}$$

$$p_{loss_l}^{Max} \geq p_{loss_l} \geq \frac{p_{s_l}^2}{V_{s_l}} R_l, \ \forall l \in l_{DCmono} \tag{4p}$$

$$p_{loss_l}^{Max} \geq p_{loss_l} \geq \frac{p_{s_l}^2}{4V_{s_l}} R_l, \ \forall l \in l_{DCbi} \tag{4q}$$

where $q_{loss_l}^{Max}$ is the upper bound of reactive power loss for AC line. $p_{loss_l}^{Max}$ is the upper bound of active power loss. These bounds are security constraints. These inequalities are now in the form of rotated second-order cone [55], [56] which is convex. Using the approximation $V_{s_l} \approx 1$, if $K_l$ denotes the squared operation capacity of line $l$, $q_{loss_l}^{Max}, p_{loss_l}^{Max}$ are determined by:

$$q_{loss_l}^{Max} = \frac{K_l}{V_{s_l}} X_l \approx K_l X_l, \ \forall l \in l_{AC} \tag{4r}$$

$$p_{loss_l}^{Max} = \frac{K_l}{V_{s_l}} R_l \approx K_l R_l, \ \forall l \in l_{DCmono} \tag{4s}$$

$$p_{loss_l}^{Max} = \frac{K_l}{4V_{s_l}} R_l \approx \frac{K_l}{4} R_l, \ \forall l \in l_{DCbi} \tag{4t}$$

At this stage, all constraints derived for the hybrid AC-DC network and FACTS devices are either linear or conic constraints. With this formulation the power flow constraints have been convexified. We denote this convex AC OPF model as SOC-ACOPF.

## IV. Wind Power Scenario Generation

In this section we propose a methodology to generate wind power scenarios based on on-site measurements of wind speed and two different wind turbine models. The proposed methodology is capable of generating scenarios for multiple wind farms, possibly owned by different companies, with any number of wind turbines. The flow chart of the wind power scenario generation methodology is illustrated by Fig. 7.

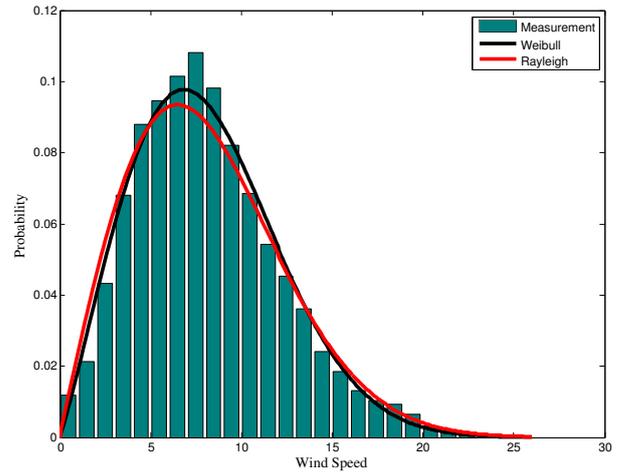

Fig. 8. The Weibull and Rayleigh distributions of the wind speed measurements.

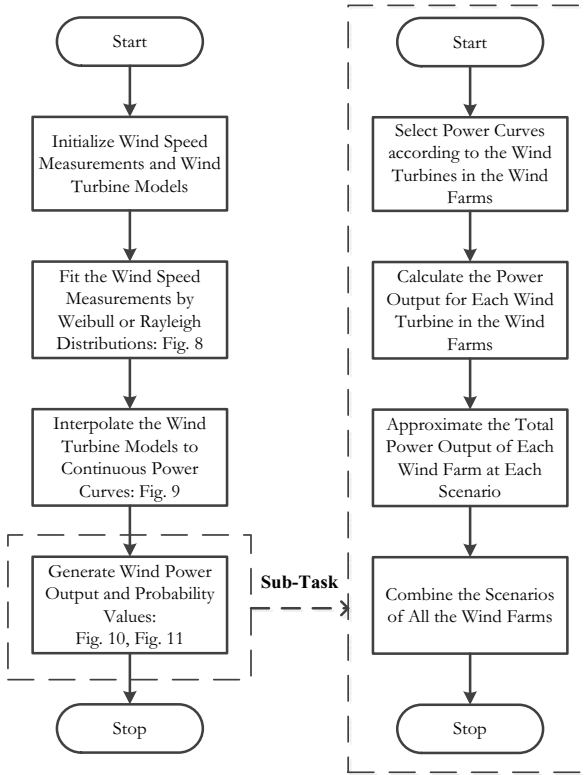

Fig. 7. The flow chart of the wind power scenario generation methodology.

The first step of our wind power scenario generation methodology consists of adequately modeling the probability distributions of the wind speed measurements. Both Weibull and Rayleigh distributions can be used to fit wind speed measurements data. In this paper, we use the wind speed measurements from Näsudden in Sweden and fit one Weibull and one Rayleigh distribution $\pi(U)$ to the measurements data where $U$ represents wind speed. The wind speed measurements and fitted distributions are demonstrated in Fig. 8. In the second step of our methodology we obtain the power output of wind turbines depending on the wind speed. In particular, we consider two different wind turbine models: the VESTAS-V90/3.0 [57] and the MERVENTO-3.6/118 [58]. For both wind turbine models there exist available power curve data sheets that link wind speed to power output at discrete given wind speeds. In order to generalize this information to be able to handle any possible wind speed $U$, we interpolate these between the existing points, which yield continuous curves $f(U)$ that relate wind speed U to power output $P$. These curves are shown in Fig. 10. More generally, with $t$ being the index for wind turbines and $j$ representing the scenario $P(t, j) = f(U_{t,j})$.

Third, we want to obtain the probability of each wind power scenario, but in order to do so let us briefly discuss the notation. We consider different wind farms $e \in E$. Let $(t \in e)$ represents the wind turbines that belong to wind farm $e$. We assume the most general case i.e. wind farms are independent of each other. This could be interpreted as each wind farm is owned by an individual company. If the wind farms were all owned by the same company our methodology still holds, however, the case would be less complex. Each wind farm (or company) $e$ considers a certain set of wind power scenarios $j_e \in J_e$. Therefore, the total number of scenarios that has to be considered is the combination of all the wind power scenarios of all the independent wind farms (companies), i.e., the cardinality of $J$: $card(J) = \prod_e card(J_e)$. As an example, suppose we have four wind farms, and each of them considers three different power output scenarios. Then, the total number of different scenarios/combinations $j$ of the power outputs of all wind farms is $3^4 = 81 (j = 1, 2, ..., 81$ in this case), as shown in Fig. 11. Scenario $j = 1$ corresponds to the case where each wind farm considers its first power output scenario. In scenario $j = 2$ the first three wind farms consider their first power output scenario, and the last wind farm considers its second power output scenario, etc. The scenario tree illustration is shown in Fig. 9. We distinguish wind power output scenarios from different wind farms by different colors. Each wind power output scenario is denoted by corresponding text. For example the first scenario from the second wind farm is denoted as F2_1 in Fig. 9. In the following equation, let $j_e(j)$ be the power output scenario of wind farm $e$ that corresponds to combination/scenario $j$. Then the probability $\pi_j$ of each scenario can be calculated as (5). The probabilities

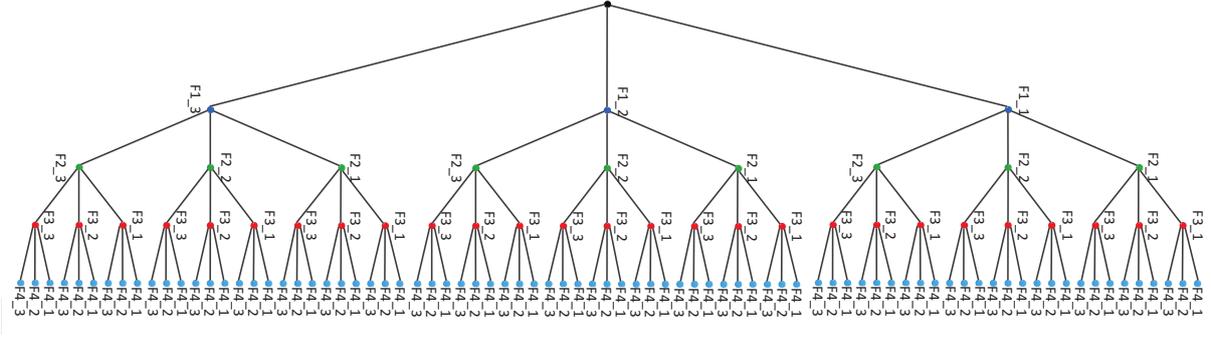

Fig. 9. The scenario tree representation of the 81 wind power scenarios.

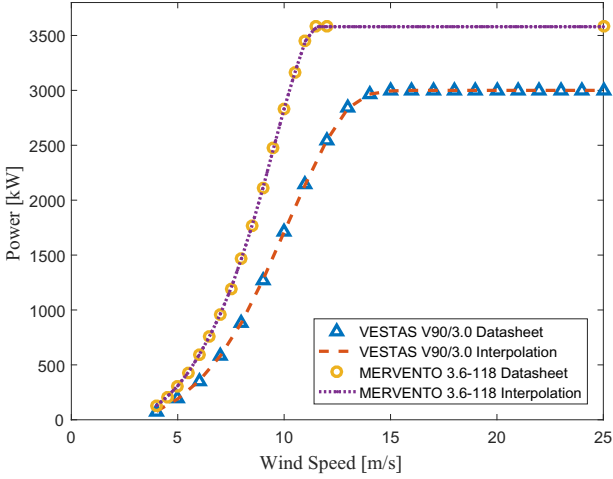

Fig. 10. The power curves of the two wind turbine models.

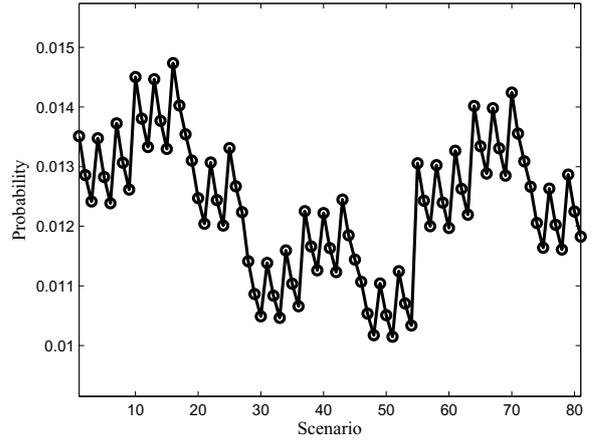

Fig. 12. The probability values of the 81 wind power scenarios

of the example of 81 scenarios are given in Fig. 12.

$$\pi_j = \prod_e \frac{\sum_{t \in e} \pi(U_{t,j_{e(j)}})}{\sum_{t \in e, j_e \in J_e} \pi(U_{t,j_{e(j)}})} \quad (5)$$

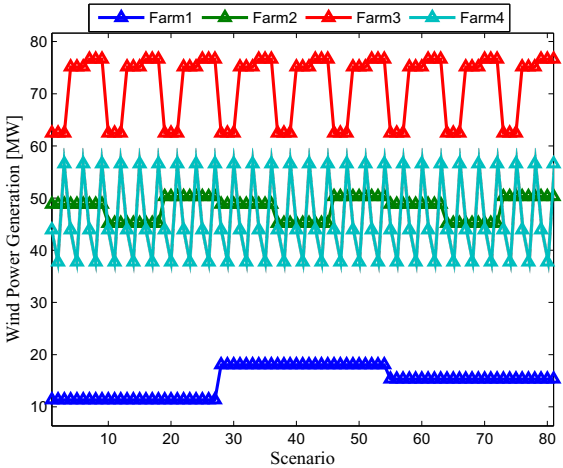

Fig. 11. The power outputs of the 81 wind power scenarios

Note that wind speed $U_{t,j_e}$ is generated from the fitted distributions in the first step of the methodology. Then $P(t, j_e)$ is calculated according to the power output curves in the second step of the methodology. The probabilities of each scenario are calculated in the third step of the methodology. Finally, we assume that wake effects and other losses of all wind farms are 15% [59], which yields the total maximum wind power generation of each wind farm $e$ in scenario $j$ as $0.85 \sum_{t \in e} P(t, j_e)$, which is the parameter $p_{e_{i,j}}^{Max}$ (maximum active power generation from wind farms) used in model (6a)-(6u). The parameter $q_{e_{i,j}}^{Max}$ (maximum reactive power generation from wind farms) in model (6a)-(6u) is generated by specifying the range of power factors of the wind farms (linear relationship between $q_{e_{i,j}}^{Max}$ and $p_{e_{i,j}}^{Max}$ is settled by the wind farms).

Finally, it is important to point out that when the number of wind farms increases, the number of wind power scenarios will increase exponentially. To deal with this challenge, we propose the M-BDA and parallel computation technique explained in Section VI. Another approach to deal with large number of wind power scenarios is scenario-reduction such as moment matching [60]. However, the approximations from scenario-reduction can lead to severe feasibility problem for the SOC-ACOPF model. One major point is, if the original number of wind power scenarios are very large, the final number of wind power scenarios after scenario-reduction (taken the approximation error tolerance into account) can still be large. So solving the challenge of large number wind power scenarios is inevitable.

## V. Stochastic SOC-ACOPF Model

We formulate the problem of optimal operation of VSC-MTDC system and FACTS as a two stage stochastic programming problem in (6a)-(6u). We assume the wind power forecast is scenario-based which is a widely used approach to consider uncertainties. The objective is to minimize the expected dispatch cost $\sum_{i \in i_{AC}} C(p_i) + \sum_{j \in J, e \in E} \pi_j C_{e,j}(p_{e_{i,j}})$ in which $\sum_{i \in i_{AC}} C(p_i)$ is the dispatch cost of thermal generators and $\sum_{j \in J, e \in E, i \in i_{AC}} \pi_j C_{e,j}(p_{e_{i,j}})$ is the expected dispatch cost of all wind farms (generally all thermal generators and wind turbines are connected to AC buses). The dispatch decisions include the power generated by both thermal generators and wind farms. The first-stage decision is to dispatch the active power generations of thermal generators $\Omega_1 = \{p_i\} \in \Re$. The solution of $\Omega_1$ is feasible for all wind power scenarios because thermal generators are dispatched before the realization of wind power. Because the reactive power generations $q_{i,j}$ of thermal generators can be adjusted rapidly according to each scenario, these variables are considered as the second-stage variables. The wind power generations $p_{e_{i,j}}, q_{e_{i,j}}$ (active and reactive power), operation points of VSC-MTDC system and FACTS constitute the set of second-stage decision variables as $\Omega_2 = \{q_{i,j}, p_{e_{i,j}}, q_{e_{i,j}}, p_{s_{l,j}}, q_{s_{l,j}}, \forall l \in l_{VSC} \cup l_{PC} \cup l_{SVC}\} \in \Re$. Where $e \in E$ is the index of wind farms, $J$ is the set of wind power generation scenarios. The set $\Omega_2$ is solved according to each wind power scenario $j$ with probability $\pi_j$. Since we have also considered the operational cost $C_{e,j}$ of the wind farms, not all wind power $p_{e_{i,j}}^{Max}$ is dispatched in the final decisions made by the operator. The amount of dispatched wind power is constrained by network conditions represented by (6b)-(6u). Note that the wind farms are connected to the power network through the VSC-MTDC system. The power balance constraints (6b)-(6c) are also valid for these connections. Because the transmission line power flow variables $p_{s_{l,j}}, q_{s_{l,j}}$ are also scenario-based, the feasibility of power balance constraints (6b)-(6c) can be guaranteed though the first-stage decision variables $p_i$ are fixed for all the wind power scenarios. It is worth to mention that in practical power system operations, both the first-stage decisions $\Omega_1$ and the second-stage decisions $\Omega_2$ are required to be made or implemented since these decisions are vital to keep the power balance of the power system. The final model can be stated as following:

$$\underset{\Omega=\Omega_1 \cup \Omega_2}{\text{Minimize}} \quad Cost = \sum_{i \in I} C(p_i) + \sum_{j \in J, e \in E, i \in I} \pi_j C_{e,j}(p_{e_{i,j}}) \tag{6a}$$

subject to

$$p_i + p_{e_{i,j}} - D_i = \sum_{l \in l_{AC}} (A_{i,l}^+ p_{s_{l,j}} - A_{i,l}^- p_{loss_{l,j}}) + G_i V_{i,j}, \\ \forall i \in i_{AC}, j \in J \tag{6b}$$

$$q_{i,j} + q_{e_{i,j}} - Q_i = \sum_{l \in l_{AC}} (A_{i,l}^+ q_{s_{l,j}} - A_{i,l}^- q_{loss_{l,j}}) - B_i V_{i,j}, \\ \forall i \in i_{AC}, j \in J \tag{6c}$$

$$q_{loss_l}^{Max} \geq q_{loss_{l,j}} \geq \frac{p_{s_{l,j}}^2 + q_{s_{l,j}}^2}{V_{s_{l,j}}} X_l, \ \forall l \in l_{AC}, j \in J \tag{6d}$$

$$p_{loss_{l,j}} X_l = q_{loss_{l,j}} R_l, \ \forall l \in l_{AC}, j \in J \tag{6e}$$

$$V_{s_{l,j}} - V_{r_{l,j}} = 2R_l p_{s_{l,j}} + 2X_l q_{s_{l,j}} - R_l p_{loss_{l,j}} - X_l q_{loss_{l,j}}, \\ \forall l \in l_{AC}, j \in J \tag{6f}$$

$$\theta_{l,j} = X_l p_{s_{l,j}} - R_l q_{s_{l,j}}, \ \forall l \in l_{AC}, j \in J \tag{6g}$$

$$p_i = \sum_{l \in l_{DC}} (A_{i,l}^+ p_{s_{l,j}} - A_{i,l}^- p_{loss_{l,j}}) + G_i V_{i,j}, \ \forall i \in i_{DC}, j \in J \tag{6h}$$

$$p_{loss_l}^{Max} \geq p_{loss_{l,j}} \geq \frac{p_{s_{l,j}}^2}{V_{s_{l,j}}} R_l, \ \forall l \in l_{DCmono}, j \in J \tag{6i}$$

$$p_{loss_l}^{Max} \geq p_{loss_{l,j}} \geq \frac{p_{s_{l,j}}^2}{4V_{s_{l,j}}} R_l, \ \forall l \in l_{DCbi}, j \in J \tag{6j}$$

$$V_{s_{l,j}} - V_{r_{l,j}} = 2p_{s_{l,j}} R_l - p_{loss_{l,j}} R_l, \ \forall l \in l_{DCmono}, j \in J \tag{6k}$$

$$V_{s_{l,j}} - V_{r_{l,j}} = p_{s_{l,j}} R_l - p_{loss_{l,j}} R_l, \ \forall l \in l_{DCbi}, j \in J \tag{6l}$$

$$-B_l^{Max} V_{s_{l,j}} \leq q_{s_{l,j}} \leq -B_l^{Min} V_{s_{l,j}}, \ \forall l \in l_{SVC}, j \in J \tag{6m}$$

$$p_{loss,Csh,j} = \frac{V_{i,j}}{R_{Csh}}, \ \forall i \in i_{DC}, j \in J \tag{6n}$$

$$p_{loss,sw,j} = \frac{V_{i,j}}{R_{sw}}, \ \forall i \in i_{DC}, j \in J \tag{6o}$$

$$\frac{8V_{i,j}}{M_{i,i'}^{Max}} \leq V_{i',j} \leq \frac{8V_{i,j}}{M_{i,i'}^{Min}}, \ \forall i \in i_{PC}, \forall i' \in i_{DC}, j \in J \tag{6p}$$

$$V_i^{Min} \leq V_{i,j} \leq V_i^{Max}, \ \forall i \in i_{AC} \cup i_{DC} \cup i_{PC}, j \in J \tag{6q}$$

$$q_{e_{i,j}}^{Min} \leq q_{e_{i,j}} \leq q_{e_{i,j}}^{Max}, \ j \in J, e \in E \tag{6r}$$

$$p_{e_{i,j}}^{Min} \leq p_{e_{i,j}} \leq p_{e_{i,j}}^{Max}, \ j \in J, e \in E \tag{6s}$$

$$q_i^{Min} \leq q_i \leq q_i^{Max}, \ \forall i \in i_{AC} \cup i_{DC} \tag{6t}$$

$$p_i^{Min} \leq p_i \leq p_i^{Max}, \ \forall i \in i_{AC} \cup i_{DC} \tag{6u}$$

Parameters $\pi_j, p_{e_{i,j}}^{Max}, p_{e_{i,j}}^{Min}, q_{e_{i,j}}^{Max}, q_{e_{i,j}}^{Min}$ are generated in the wind power scenario generation methodology explained in Section IV. The optimization problem (6a)-(6u) minimizes the expected power generation cost considering the uncertainties of wind power. To show the size of the stochastic SOC-ACOPF model, the number of decision variables $N_{var}$ in the model is expressed as:

$$N_{var} = \bar{J} \left[5\bar{L} + \bar{G} + 2\bar{E} + \bar{I}\right] + \bar{G} \tag{7a}$$

where $\bar{J}, \bar{L}, \bar{G}, \bar{E}, \bar{I}$ are the cardinalities of the sets $J, L, G, E, I$ correspondingly. Especially, $G$ is the set of generators (note not all nodes in the network are connected with generators). The number of constraints $N_{con}$ in the model is:

$$N_{con} = \bar{J}[2\bar{i_{AC}} + 5\bar{l_{AC}} + \bar{i_{DC}} + 5\bar{l_{DC}} + \bar{Csh} + \bar{sw} + \\ 2\bar{CONV} + 2\bar{l_{svc}} + 2\bar{E} + 2\bar{I}] + 2\bar{G} \tag{7b}$$

Similarly, the $\bar{i_{AC}}, \bar{l_{AC}}, \bar{i_{DC}}, \bar{l_{DC}}, \bar{Csh}, \bar{sw}, \bar{CONV}, \bar{l_{svc}}$ are the cardinalities of the corresponding sets $i_{AC}, l_{AC}, i_{DC}, l_{DC}, Csh, sw, CONV, l_{svc}$. Especially, $Csh$ is the set of converter shunt resistor, $sw$ is the set of switching loss resistor, $CONV$ is the set of voltage converters (see the model details in Section III). Same expressions hold for sizes of variables and constraints by using the nonconvex AC OPF model to formulate the stochastic problem. The only difference is we replace the voltage variable $v_i$ in the nonconvex AC OPF model by the voltage square variable $V_i$ in our SOC-ACOPF model. Besides, each nonconvex constraint in the nonconvex AC OPF model is replaced by the corresponding convex formulation in our SOC-ACOPF model. Please note the bounds for the corresponding variables in both nonconvex ACOPF and SOC-ACOPF models are the same.

If we use a convex (generally quadratic) cost function for $Cost$, because both the objective and constraints are convex, this minimization problem is convex. This is further validated by the numerical results in Section VII of this paper through using MOSEK solver in GAMS which can only solve convex optimization problems. Hence, the global optimal solutions can be obtained by solving (6a)-(6u). The challenge is, when large number of wind power scenarios are considered, the scale of the optimization problem (6a)-(6u) can exceed the capability of the optimization solver. We explain in the next section of this paper how to deal with this challenge by decomposition and parallelization.

## VI. Decomposition of Stochastic SOC-ACOPF

If we assume the OPF problem is feasible (at least one solution exists), then strong duality holds for the formulated SOC-ACOPF model. Thus Benders decomposition can be applied to solve the large-scale stochastic SOC-ACOPF problem in a decomposed way [61]. In Benders decomposition, the original large-scale optimization problem is decomposed into a master problem and several subproblems. The objective solution of the master problem gives a lower bound on the optimal objective solution of the original minimization problem. The subproblems can give upper bound of the objective in the original minimization problem. Over the course of the iterations, the lower and upper bounds given by the master problem and subproblems gradually converge. The challenges are how to formulate the master problem and subproblems in the decomposition, and how to accelerate the convergence. These challenges are addressed in this section.

Taking the first stage decision variables $\Omega_1 = \{p_i\} \in \Re$ as the complicating variables in Benders decomposition, we can decompose the stochastic SOC-ACOPF (6a)-(6u). The master problem of the proposed M-BDA is formulated as:

$$\underset{\Omega_1}{\text{Minimize}} \quad Cost = \sum_{i \in i_{AC}} C(p_i) + Cost_w \qquad (8a)$$

subject to $(6u)$

$$Cost_w \geq Cost_{w,k} + \sum_{i \in i_{AC}} \mu_{i,k}^p (p_i - \hat{p}_{i,k}), \ \forall k \in K \qquad (8b)$$

where $\hat{p}_{i,k}$ is the dispatched active power (first stage decision) at iteration $k$. Constraint (8b) is the Benders cut which expands iteratively in order to take more information from the solutions of the subproblems into account (a new Benders cut is added from each subproblem to the master problem during each iteration). $Cost_{w,k} = \sum_n Cost_w^n$ is the sum of objective solutions in the subproblems at iteration $k$. $Cost_w^n$ is the objective of subproblem $n$ (the expected dispatch cost of wind power generations). $\mu_{i,k}^p$ is the dual variable associated with (9b) at iteration $k$ (equal to zero in the first iteration) in the subproblem. The subproblem of the proposed M-BDA is:

$$\underset{\Omega_2}{\text{Minimize}} \quad Cost_w^n = \sum_{j \in j_n, e \in E, i \in i_{AC}} \pi_j C_{e,j}(p_{e_{i,j}}) \qquad (9a)$$

subject to $(6b) - (6t)$

$$p_i = \hat{p}_{i,k} : \mu_{i,k}^p, \ \forall k \in K \qquad (9b)$$

where $\cup_n^T j_n = J$ is a division of the wind power scenario set $J$. $N = \sum n$ is the total number of threads settled in the parallel computation. The scenarios assigned to subproblem $n$ are included in the set $j_n$. To guarantee the feasibility of all the subproblems in M-BDA, we allow load shedding for all the buses in the network. We set the cost parameters of load shedding much larger than the most expensive generators in the network. Although some loads may not be fully met in the initial iterations, the final solutions of M-BDA cover all these loads (because the cost of load shedding is too expensive and will be iteratively met by power generation in the original network). Our simulations show that this method is more efficient to guarantee the feasibility of the subproblems than using the method of the original feasibility cut in Benders decomposition (the MOSEK solver can not converge after several hours when using the original feasibility cut approach).

As the iterations proceed, more Benders cuts are added to the master problem (each subproblem submits its own Benders cuts. see Fig. 13). After solving the master problem (8), all subproblems can be solved in parallel. The proposed parallel computation structure of stochastic SOC-ACOPF using the proposed M-BDA is illustrated in Fig. 13. The Master problem is responsible to give solutions of the first-stage decision variables $\Omega_1 = \{p_i\} \in \Re$ while the subproblems yield solutions of the second-stage decision variables $\Omega_2 = \{q_{i,j}, p_{e_{i,j}}, q_{e_{i,j}}, p_{s_{l,j}}, q_{s_{l,j}}, \forall l \in l_{VSC} \cup l_{PC} \cup l_{SVC}\} \in \Re$. There is no communication requirement between the subproblems. There are three major modifications we have made in the formulation the proposed M-BDA compared with the original Benders decomposition method: (1) We take the active power generation of conventional power generators as the complicating variable to formulate the master problem and subproblems of M-BDA; (2) Instead of formulating one single sub-problem, we decompose the wind power dispatch problem based on the wind power scenarios and formulate multiple subproblems. In this way, the scale or complexity of each subproblem is largely reduced to reduce the computational burden. The number of wind power scenarios assigned to each subproblem can be balanced to facilitate parallel computing; (3) We guarantee the feasibility of the formulated subproblems during the iterations of M-BDA by allowing load shedding. This approach avoids using the original Benders feasibility cut approach which shows less efficiency in our computations.

## VII. Numerical Results

### A. Comparison of SOC-ACOPF model and DC OPF model

Because of the relaxations and approximations used to convexify the non-convex constraints of the AC network, VSC-MTDC system and FACTS devices, the solutions obtained from the proposed SOC-ACOPF model are approximated solutions. The benefits of using the SOC-ACOPF include proved guarantee of global optimal solution and the convergence of M-BDA explained in section VI. Because we approximate fewer constraints compared with the widely used DC OPF model, the feasibility (satisfying the original non-convex AC OPF constraints) of SOC-ACOPF model is better than the DC OPF model. Considering the enormous applications of DC OPF in operating power system, the SOC-ACOPF can also serve as an improved tool for power system operations. To give an overall comparison of SOC-ACOPF and DC OPF, we calculate the difference between the left side and right side of AC constraints (equations) as the AC feasibility gap and list the maximum absolute values of AC feasibility gaps of the proposed SOC-ACOPF model and DC OPF model in Table I. In all the test cases, the AC feasibility of the proposed SOC-ACOPF model is much better than the DC OPF model for the power balance constraints (1a)-(1b) and power loss constraints (1c)-(1d). For the voltage drop constraints (1e)-(1f), there is no significant AC feasibility difference between the SOC-ACOPF model and the DC OPF model. In summary, AC feasibility of the proposed SOC-ACOPF model is better than the DC OPF model. The computation CPU time results are listed in Table II. The SOC-ACOPF model requires a bit more computation time than the DC OPF model to find more accurate solutions. This is reasonable since the SOC-ACOPF model has more constraints than the DC OPF model.

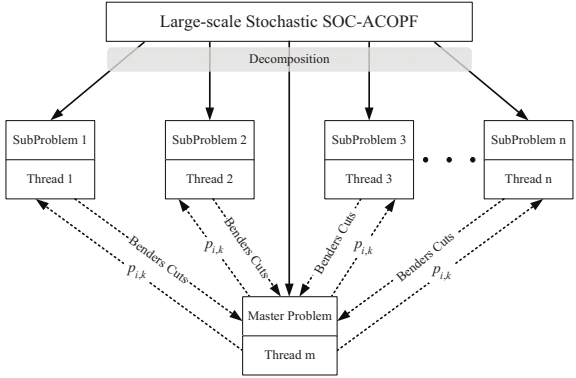

Fig. 13. The proposed M-BDA to solve the stochastic SOC-ACOPF by parallel computation.

### B. Solve stochastic SOC-ACOPF by M-BDA and parallel computation

The test case network configuration is shown in Fig. 14. Two wind farms are connected to the IEEE30-Bus network through a five-terminal VSC-MTDC system (D1-D5). The VSC-MTDC system is connected to bus 1, bus 15 and bus 30 in the IEEE30-Bus network. There are twenty VESTAS-V90/3.0 wind turbines in the first wind farm and thirty MERVENTO-3.6/118 wind turbines in the second wind farm. Two STATCOM devices (located at bus 4 and bus 18) and two SVC devices (located at bus 14 and bus 21) are connected in the AC network. The proposed wind power scenario generation methodology is implemented in MATLAB. The stochastic SOC-ACOPF model and M-BDA are coded in GAMS and solved by MOSEK [62]. The test cases with 10, 50, 100, 500, 1000 and 5000 wind power scenarios are run on a PC with Intel i7-2760QM 2.4 GHz CPU and 8 GB of RAM. Simulations of the 10000 and 50000 wind power scenarios test case are performed at PDC Center for High Performance Computing (PDC-HPC) in KTH Royal Institute of Technology. The deployed computation node at PDC-HPC has 48 CPU cores (3.00 GHz Intel Xeon Processor E7-8857) and 1 TB of RAM in total [63].

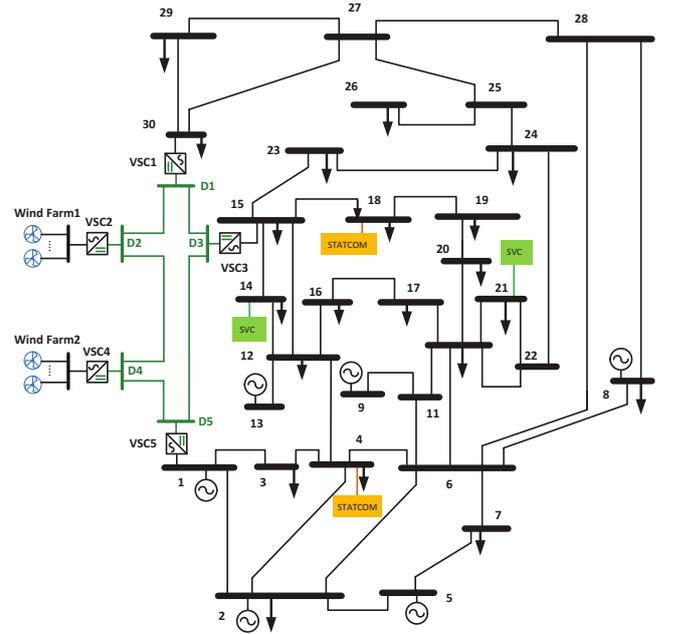

Fig. 14. The IEEE30-Bus network with the two wind farms integrated by the VSC-MTDC system.

In order to compare the performance of different solution approaches, the following solution approaches are defined:

1) **Single-Stage** solution approach. The stochastic programming for all wind power scenarios are formulated and solved as one single stochastic optimization problem. This approach is set as the benchmark to compare with the proposed M-BDA approach.

2) **Parallel M-BDA** solution approach. The subproblems in M-BDA are solved in parallel during each iteration. This approach is proposed to improve the computational efficiency when large number of wind power scenarios are considered.

3) **Serial M-BDA** solution approach. The subproblems in M-BDA are solved in serial during each iteration. This solution approach is to demonstrate the efficiency of the proposed parallel computation structure in GAMS.

The proposed M-BDA stops when the upper and lower bounds of the objective is within 2%. The solution approaches are illustrated in Fig. 15. In Fig. 15, we assign two wind power scenarios to each subproblem for ease of illustration. In the computation of a large number of wind power scenarios, we can assign more scenarios to each subproblem to distribute the computation burden to each thread more evenly. The numerical results of objective values are summarized in Table III. We can see that solutions by M-BDA are all very close (relative error less than 2%) to the single-stage solution. The objective solutions are different for different test cases because the generated wind power scenarios (probability values and maximum power outputs) are different. When considering 10000 wind power scenarios, MOSEK cannot converge using the single-stage approach. The proposed M-BDA shows strong convergence capability and finds the optimal solutions for all the test cases.

TABLE I
Summary of the Maximum Absolute Values of the AC Feasibility Gap

| Test Case | Model | Feasibility Gap of the AC Constraints | | | | | |
|---|---|---|---|---|---|---|---|
| | | (1a) | (1b) | (1e) | (1f) | (1c) | (1d) |
| IEEE14 | SOC-ACOPF | 1.17E-09 | 2.95E-09 | 7.73E-02 | 1.03E-02 | 1.12E-09 | 5.51E-10 |
| | DC OPF | 1.42E-09 | 2.47E-03 | 8.80E-02 | 4.87E-03 | 1.35E-01 | 4.43E-02 |
| IEEE57 | SOC-ACOPF | 1.00E-08 | 8.59E-09 | 1.26E-01 | 5.35E-03 | 8.30E-02 | 2.91E-09 |
| | DC OPF | 1.68E-09 | 7.59E-03 | 1.11E-01 | 8.90E-03 | 2.34E-01 | 4.59E-02 |
| IEEE118 | SOC-ACOPF | 2.89E-09 | 1.00E-08 | 7.48E-02 | 2.07E-02 | 5.65E-01 | 2.87E-10 |
| | DC OPF | 2.00E-08 | 3.51E-02 | 7.75E-02 | 1.12E-02 | 5.00E-01 | 5.88E-02 |
| IEEE300 | SOC-ACOPF | 2.00E-08 | 2.00E-08 | 2.72E-01 | 2.57E-02 | 7.40E-01 | 1.70E-09 |
| | DC OPF | 4.86E-05 | 2.43E-01 | 1.82E-01 | 2.97E-02 | 3.34E+00 | 2.09E-01 |
| 1354pegase | SOC-ACOPF | 3.30E-07 | 3.11E-05 | 1.70E-01 | 5.34E-02 | 8.78E-01 | 1.32E-02 |
| | DC OPF | 7.00E-07 | 1.95E-01 | 2.31E-01 | 1.88E-02 | 2.73E+00 | 2.07E-01 |
| 2869pegase | SOC-ACOPF | 5.08E-06 | 1.76E-05 | 2.89E-01 | 7.63E-02 | 2.26E+00 | 1.72E-02 |
| | DC OPF | 4.01E-04 | 2.61E-01 | 2.32E-01 | 2.41E-02 | 2.10E+00 | 1.61E-01 |

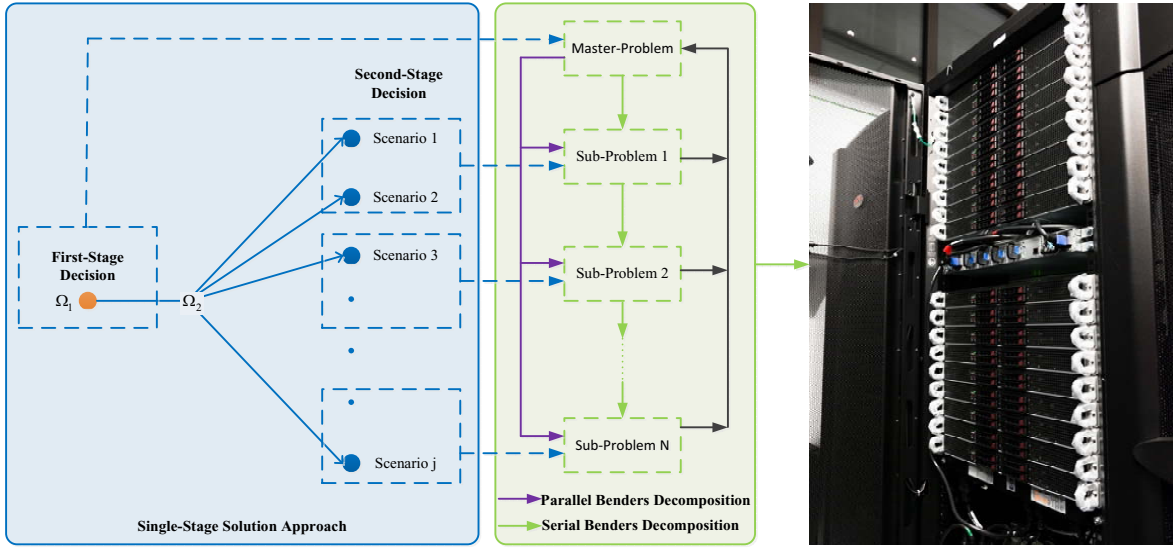

Fig. 15. Solution Approaches of the Stochastic SOC-ACOPF Model.

TABLE II
Summary of the Computation CPU Time

| Test Case | SOC-ACOPF | DC OPF |
|---|---|---|
| IEEE14 | 0.08 | 0.03 |
| IEEE57 | 0.09 | 0.03 |
| IEEE118 | 0.09 | 0.08 |
| IEEE300 | 0.25 | 0.09 |
| 1354pegase | 0.76 | 0.26 |
| 2869pegase | 1.97 | 1.37 |

We list the CPU computation time for the single-stage and M-BDA approaches in Table IV. For the test cases of 10, 50, 100, 500 and 1000 wind power scenarios, we set 4 threads for the parallel computation of M-BDA. 8 threads are set for the parallel computation of 5000 wind power scenarios. 30 threads are set for the parallel computation of 10000 scenarios test case. We set 50 threads for the parallel computation of 50000 wind power scenarios. In all the test cases, parallel M-BDA is faster than the serial execution. In the test cases of 10000 and 50000 wind power scenarios, parallel M-BDA outperforms the single-stage approach in terms of both computation speed and convergence capability. For the test case of 10000 wind power scenarios, the single-stage approach cannot converge within 184.27 seconds. For the test case of 50000 wind power scenarios, the single-stage approach fails to solve the Stochastic SOC-ACOPF model after over 4 hours CPU time. As a comparison, HPC executed M-BDA solves the stochastic SOC-ACOPF model with 50000 wind power scenarios within 23 minutes CPU time. Even though for small numerical examples the single-stage solution approach is faster than M-BDA, it is important to point out that for large-scale cases, the single-stage solution approach cannot converge due to the curse of dimensionality. The proposed M-BDA, however, finds the global optimal solution faster than it takes for MOSEK to determine that if there is a convergence problem for the single-stage solution approach. Therefore, our solution approach is useful for large-scale computation. It is worth to mention that even though we use a small power network in which there are 30 buses, the total number of wind power scenarios that we have considered here show that the scale of the stochastic SOC-ACOPF model is very large. This is demonstrated by the total number of constraints of each test case in Table IV. Note that the total number of constraints in Table IV refers to the coded model in GAMS which, in general, has more constraints than the mathematical model represented in (6a)-(6u).

TABLE III
OBJECTIVE VALUE

| Number of Scenarios | Single-Stage[$] | M-BDA | | |
|---|---|---|---|---|
| | | Number of Threads | Upper Bound[$] | Lower Bound[$] |
| 10 | 320.26 | 4 | 325.01 | 319.69 |
| 50 | 354.91 | 4 | 356.00 | 349.99 |
| 100 | 320.06 | 4 | 324.33 | 319.87 |
| 500 | 383.04 | 4 | 386.13 | 383.44 |
| 1000 | 380.65 | 4 | 382.77 | 380.46 |
| 5000 | 380.19 | 8 | 383.33 | 380.05 |
| 10000 | NA | 30 | 378.55 | 371.67 |
| 50000 | NA | 50 | 392.03 | 391.16 |

TABLE IV
CPU COMPUTATION TIME

| Number of Scenarios | Number of Constraints | Single-Stage [s] | M-BDA [s] | |
|---|---|---|---|---|
| | | | Serial | Parallel |
| 10 | 4560 | 0.23 | 1.94 | 0.59 |
| 50 | 22680 | 0.72 | 8.585 | 1.841 |
| 100 | 45330 | 1.72 | 11.12 | 3.86 |
| 500 | 226530 | 7.66 | 143.54 | 39.61 |
| 1000 | 453030 | 28.74 | 296.98 | 70.87 |
| 5000 | 2265030 | 104.05 | 3387.69 | 550.89 |
| 10000 | 4530030 | 184.27 | 6191.58 | 116.34 |
| 50000 | 22650030 | 15336.25 | 54726.86 | 1343.32 |

The convergence of the proposed M-BDA is shown in Fig. 16. Note the Y-axis in this figure is in base-10 logarithmic scale. When the number of wind power scenarios increases, the required number of iterations of M-BDA also increases. The efficiency of the proposed M-BDA can be observed in that even in the case of 50000 scenarios, the M-BDA converges within 14 iterations.

### C. Value of Stochastic Solution

To show the benefits of the stochastic programming, we calculate the Value of Stochastic Solution ($VSS$) according to [64] as:

$$VSS = Cost^{D*} - Cost^{S*} \qquad (10a)$$

Where $Cost^{D*}$ is the objective value of deterministic solution, $Cost^{S*}$ is the objective value of stochastic solution. $VSS$ is a measure of the benefits to model the uncertain power outputs of wind farms as scenario-based variables instead of using the expected values. $Cost^{D*}$ is obtained by firstly replacing

the wind power maximum generation parameters $p_{e_{i,j}}^{Max}, q_{e_{i,j}}^{Max}$ with the expected values:

$$p_{e_{i,j_0}}^{Max} = \sum_j \pi_j p_{e_{i,j}}^{Max} \qquad (10b)$$

$$q_{e_{i,j_0}}^{Max} = \sum_j \pi_j q_{e_{i,j}}^{Max} \qquad (10c)$$

Then we solve the model (6a)-(6u) (in this way there is only one scenario $j = j_0$ in the model) to find the deterministic solution of the first-stage decision variables $p_i^{D*}$. After obtaining $p_i^{D*}$, we fix the first-stage decision variables as:

$$p_i = p_i^{D*} \qquad (10d)$$

Afterwards, we solve the stochastic model (6a)-(6u) again to find $Cost^{D*}$. Note, in this step, the second-stage parameters $p_{e_{i,j}}^{Max}, q_{e_{i,j}}^{Max}$ are not fixed as expressed in (10b)-(10c) and any number of wind power scenarios are allowed.

The results of $VSS$ are listed in Table V. Note the $VSS$ for the test case of 10000 and 50000 wind power scenarios are obtained by using the proposed M-BDA approach. Significant benefits of stochastic programming are observed when large number of wind power scenarios are considered. The reason for large value of $VSS$ in cases having large number of wind power scenarios is because the value of lost load (VoLL) parameter is large in our test cases. These results show that, when large number of wind power scenarios are considered, it is very hard to find the deterministic solution $Cost^{D*}$ without load shedding. In contrast, the stochastic programming approach can always find the optimal solutions efficiently.

### D. Feasibility of Solutions from the Stochastic SOC-ACOPF Model

To demonstrate the AC feasibility (satisfying the original power network constraints) of solutions obtained from the proposed stochastic SOC-ACOPF model, we list the feasibility gap results in Table VI. All solutions show very small feasibility gap performance which means satisfying the original power network constraints well.

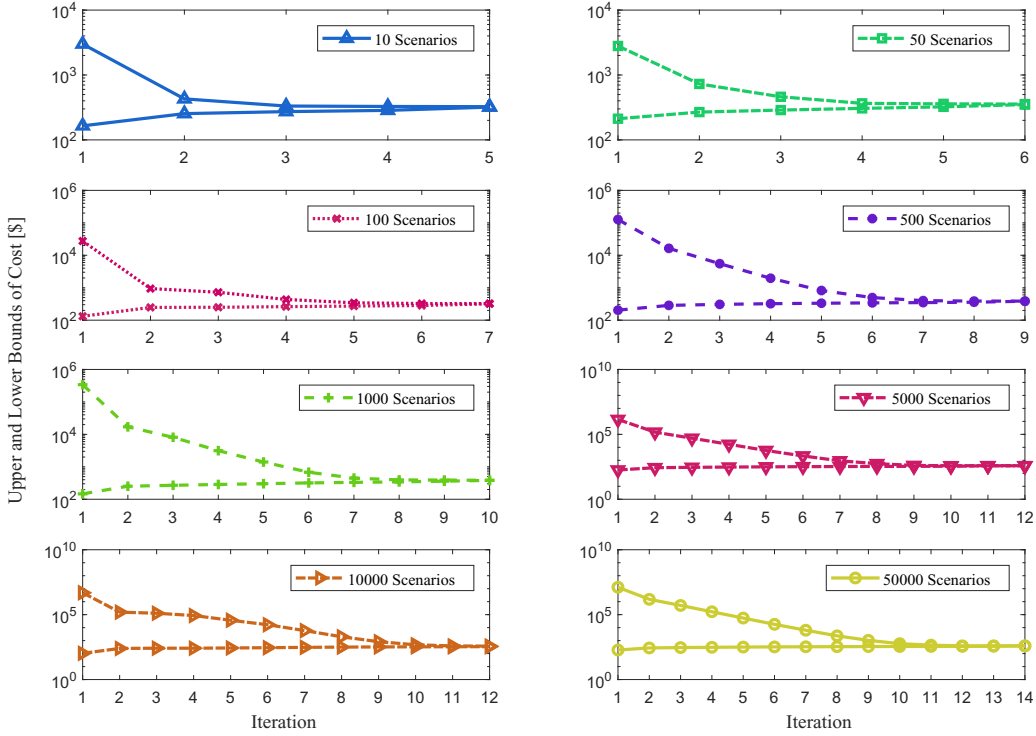

Fig. 16. The Convergence of M-BDA.

TABLE V
VALUE OF STOCHASTIC SOLUTION ($VSS$)

| Number of Scenarios | Solution [$] | | $VSS$ [$] |
| --- | --- | --- | --- |
| | Deterministic | Stochastic | |
| 10 | 478.71 | 320.26 | 158.45 |
| 50 | 537.46 | 354.91 | 182.55 |
| 100 | 532.77 | 320.06 | 212.71 |
| 500 | 7722.52 | 383.04 | 7339.48 |
| 1000 | 9924.11 | 380.65 | 9543.46 |
| 5000 | 56548.50 | 380.19 | 56168.31 |
| 10000 | 65960.41 | 378.55 | 65581.86 |
| 50000 | 88233.65 | 392.03 | 87841.62 |

TABLE VI
FEASIBILITY GAP OF SOLUTIONS FROM THE STOCHASTIC SOC-ACOPF MODEL

| Number of Scenarios | Maximum Feasibility Gap |
| --- | --- |
| 10 | 8.00E-08 |
| 50 | 1.20E-07 |
| 100 | 1.59E-05 |
| 500 | 5.39E-06 |
| 1000 | 1.00E-06 |
| 5000 | 1.02E-04 |
| 10000 | 5.49E-06 |
| 50000 | 1.50E-07 |

## VIII. CONCLUSIONS

We propose a stochastic SOC-ACOPF model to optimally operate the power network incorporating VSC-MTDC system and FACTS devices in the context of scenario-based forecast of wind power. Both mono-polar and bipolar MTDC connections are modeled. STATCOM and SVC as FACTS devices are integrated in a unified modelling approach in the proposed SOC-ACOPF model. The SOC-ACOPF is a convex optimization problem which can be solved efficiently to global optimality. Using the wind speed measurements and the wind turbine models, we are able to generate representative power generation scenarios from multiple wind farms. Finally, we propose to use M-BDA to address the computational challenge of large-scale power system operations. When 10000 and 50000 wind scenarios are considered, a single-stage stochastic SOC-ACOPF model is not tractable even by high performance computing. As a comparison, the proposed parallel computation in GAMS based on M-BDA is capable of solving the stochastic SOC-ACOPF model and accelerating the computations. Large values of $VSS$ show the significant benefits of using stochastic SOC-ACOPF when large number of wind power scenarios are considered. The results of this paper demonstrate the great potential of stochastic conic programming, high performance computing and M-BDA in dealing with the severe challenge of integrating large-scale wind power which is inevitable in the future of power system. Future work to investigate our approaches for larger power networks in the context of different types of renewable energy resources is expected.

Considering the wind power correlations between separate wind farms can also be an interesting extension of our work.

## Acknowledgment

The authors would like to thank the support from SETS Erasmus Mundus Joint Doctorate Fellowship for this research project on wind power. We want to express special thanks to Prof. Benjamin F. Hobbs from The Johns Hopkins University for insightful discussions about the methods used in this paper. Some computations in this paper are performed on resources provided by the Swedish National Infrastructure for Computing (SNIC) at PDC Centre for High Performance Computing (PDC-HPC) in KTH Royal Institute of Technology. Our great thanks also go to Jing Gong as Application Expert in PDC-HPC for helping us configure MOSEK solver in the Linux environment and other remote connection settings. All the careful reviews on this paper by the editors and reviewers of this journal are highly appreciated by the authors.

**Zhao Yuan** is an Assistant Professor and the Head of the Electrical Power Systems Laboratory (EPS-Lab) at University of Iceland. He received joint PhD degree from KTH Royal Institute of Technology, Comillas Pontifical University and Delft University of Technology in 2018. Zhao proposed and proved the theorems about the existence and uniqueness of the optimal solution of the convex optimal power flow model based on second-order cone programming. He co-developed the Energy Management System (EMS) of the 560kWh / 720kVA Battery Energy Storage System (BESS) on Swiss Federal Institute of Technology Lausanne (EPFL) campus and the Aigle Smart Grid in Switzerland in 2020. His research work on coordinated transmission-distribution operation was awarded by Advances in Engineering (AIE) as Key Scientific Article Contributing to Excellence in Science and Engineering Research in 2017. His article "Second-order cone AC optimal power flow: convex relaxations and feasible solutions" is awarded as the 2019 Best Paper in the Journal of Modern Power Systems and Clean Energy (MPCE).